\documentclass[prd,twocolumn,nofootinbib,showpacs,preprintnumbers]{revtex4}
\pdfoutput=1
\usepackage{amsfonts,amsmath,amssymb}
\usepackage{graphicx}
\usepackage{color}
\usepackage{natbib}
\usepackage{enumitem}

\usepackage[plainpages=false, colorlinks=true, anchorcolor=blue, linkcolor=blue, citecolor=blue, bookmarks=false]{hyperref}
\usepackage{multirow}
\newcommand{\rthis}[1]{\textcolor{black}{#1}}

\begin{document}
\newcommand{\apjl}{Astrophys. J. Lett.}
\newcommand{\apjs}{Astrophys. J. Suppl. Ser.}
\newcommand{\aap}{Astron. \& Astrophys.}
\newcommand{\aj}{Astron. J.}
\newcommand{\araa}{Ann. Rev. Astron. Astrophys. } 
\newcommand{\mnras}{Mon. Not. R. Astron. Soc.}
\newcommand{\apss} {Astrophys. and Space Science}
\newcommand{\jcap}{JCAP}
\newcommand{\pasj}{PASJ}
\newcommand{\pasa}{Pub. Astro. Soc. Aust.}
\newcommand{\physrep}{Physics Reports}

\title{Robust model comparison tests of DAMA/LIBRA annual modulation  }
\author{Aditi \surname{Krishak}$^1$}\altaffiliation{E-mail:aditi16@iiserb.ac.in}
\author{Aisha \surname{Dantuluri}$^2$}  
\altaffiliation{E-mail: adantulu@ucsd.edu}
\author{Shantanu  \surname{Desai}$^3$ } 
\altaffiliation{E-mail: shntn05@gmail.com}


\affiliation{$^{1}$ Department of Physics, Indian Institute of Science Education and Research, Bhopal, Madhya Pradesh 462066, India}
\affiliation{$^{2}$ Department of Mathematics, University of California, San Diego, La Jolla, CA 92093, USA}
\affiliation{$^{3}$Department of Physics, Indian Institute of Technology, Hyderabad, Telangana-502285, India}

\begin{abstract}
 We evaluate the statistical significance of the DAMA/LIBRA claims for annual modulation using three independent model comparison techniques, viz frequentist, information theory, and Bayesian analysis. We fit the data from the DAMA/LIBRA experiment to both cosine and a constant model, and carry out model comparison by choosing the constant model as the null hypothesis.  For the frequentist test, we invoke Wilk's theorem and calculate the significance using $\Delta \chi^2$ between the two models. For information theoretical tests, we calculate the difference in Akaike Information Criterion (AIC)  and Bayesian Information criterion (BIC) between the two models. We also compare the two models in a Bayesian context by calculating the Bayes factor.
 \rthis{We also search for higher harmonics in the DAMA/LIBRA data using generalized Lomb-Scargle periodogram. We finally test the sensitivity of  these model comparison techniques in discriminating between pure noise and a cosine signal using synthetic data.} This is the first proof of principles application of AIC, BIC as well as Bayes factor  to the DAMA data. All our analysis codes along with the data used in this work have been made publicly available.
\end{abstract}


\maketitle

\section{Introduction}
The dark matter problem~\cite{Kamionkowski,Silk} is one of the most important vexing problems in astrophysics eluding a solution, ever since its existence was first pointed out 80-90 years ago  by  Oort and Zwicky~\cite{Hooper}. The current concordance model of cosmology indicates that about 27\% of universe contains cold dark matter~\cite{Planck15}.  It has been known since the 1970s that any elementary particle, which is a thermal relic from the Big-Bang and with electroweak scale interactions with ordinary matter satisfies all the properties needed for a cold dark matter candidate, such as the correct  relic abundance, non-relativistic velocities at the time of decoupling, etc~\cite{Weinberg}. Therefore, such weakly interacting massive particles (hereafter, WIMPs) are the most favored   dark matter candidates. 

A whole slew of experiments and detection methods have been employed to experimentally  detect WIMPs and measure its properties~\cite{Bauer}. These include direct production of WIMPs at collider experiments, indirect signatures of WIMP annihilation products in cosmic rays, neutrinos, or photons, and direct detection of dark matter in underground cryogenic experiments. At the time of writing, there is no incontrovertible evidence for any smoking-gun signatures of WIMP annihilation  or direct production of WIMPs (see Refs.~\cite{Desai04,Cirelli,Cirelli15,LHC18,Hooper18} and references therein for most recent updates.) 

Among the plethora of direct dark matter detection experiments, only one experiment (viz. the DAMA experiment in Gran Sasso) has argued for the detection of dark matter. The observational signature found by the DAMA collaboration indicative of dark matter, is the detection of annual modulation in their residual count rates, which is based on the  vector sum  of  the motion of our Sun moving in our galaxy with respect to the Local Standard of Rest and  the Earth's revolution around the Sun. It was pointed out in the 1980s~\cite{Drukier,Freese88}, that  because of these motions,  any dark matter experiment should detect a peak flux of WIMP-induced interactions  around June 2 (when Earth's orbital velocity gets added to that of the Sun with respect to the Galaxy) and a minimum around December 2 (when Earth's orbital velocity gets subtracted compared to that of the Sun). For more than 20 years, DAMA experiment has found such  an annual modulation in their residual count rates, which has exactly the above signatures.  The first phase of DAMA started in 1995.  At the time of the first data release, the evidence was $3\sigma$~\cite{DAMA00}. The statistical significance of this annual modulation has steadily increased with accumulated data taking, during the different phases of the DAMA experiment~\citep{DAMA03,Dama08,DAMA10,Dama13,Dama18}. At the end of the first phase of the DAMA experiment (called DAMA/NaI), the statistical significance had increased to 6.3$\sigma$~\cite{DAMA03}.
In the latest data release from phase 2 of the upgraded DAMA/LIBRA experiment (where the energy threshold has been lowered), the net statistical significance in the 2-6 keV energy bin, after combining data from all the phases of the experiment   is 12.9$\sigma$ and greater than 8$\sigma$ in other energy intervals~\citep{Dama18}.

One problem with the dark matter interpretation of the above claim is that this signal has not been confirmed by any other direct detection experiment, and the entire allowed region enclosing the WIMP mass and cross-section (inferred from the DAMA annual modulation), has been ruled out by a number of other direct dark matter detection experiments~\cite{Lux,PandaX,Aprile,CDEX}. 
See for example Ref.~\cite{Rauch,Schumann} for a recent review of all direct dark matter searches.

Although a number of  particle physics, nuclear physics,  and astrophysics explanations have been concocted to reconcile  the incompatibility of DAMA results with other experiments (eg.~\citep{Freese08,Zupan,Catena, Gelmini,Williams,Tomar} and references therein), none of them can satisfactorily explain all the DAMA observations. However, until recently, none of  the other direct detection dark matter experiments had  used the same target as DAMA  (NaI) or were sensitive to the same annual modulation signature found in DAMA; but this has been recently rectified.
In 2019,  two experiments ANAIS-112 and COSINE-100, both of which use NaI(Tl) as the detector target released their first results. The ANAIS-112 experiment located in the Canfranc Underground Laboratory in Spain released results from modulation analysis of 1.5 years of data from 157.55 kg years exposure~\citep{Anais}. 
The COSINE-100 experiment~\citep{Cosine} located in the Yangyang underground laboratory in South Korea released their first results using 1.7 years of data with a total exposure of 97.7 kg years. The current exposures of these experiments is not large enough to robustly rule out the DAMA claim. This annual modulation claim will also be tested by other experiments, such as DM-Ice17~\citep{dmice}, KIMS~\cite{Kims}, and SABRE~\cite{Sabre}. 

The DAMA collaboration has calculated the statistical significance of the annual modulation claim using the frequentist test, by comparing the difference in $\chi^2$ between the null hypothesis and the sinusoidal model~\citep{Dama18}. Although a few  other groups have done an independent statistical analysis of the DAMA data~\cite{Sturrock,Freese17,Freese18,Kahlhoefer}, a model comparison analysis of the DAMA annual modulation claims using Bayesian or information theory methods has not been previously done.  It is important to test the robustness of the claim using independent model comparison techniques.
Such model comparison techniques are now routinely used in Cosmology to weigh  the evidence for $\Lambda$CDM model versus alternatives~\citep{Liddle,Liddle07,Shi,Trotta,Weller}. Previously in this field, Bayesian techniques have been used to test if the CoGeNT annual modulation data can be explained by dark matter induced scatterings~\cite{Cogent}. They have also been used to test the compatibility of upper limit on WIMP couplings from XENON-100 experiment, with the annual modulation seen in DAMA and CoGeNT~\cite{Arina} (see also Ref.~\cite{Algeri}). 
This is the main goal of this work.
The techniques used in this work can be easily applied to evaluate the significance  of  annual modulation claims of  other experiments, some of which have already produced preliminary results.



The outline of this paper is as follows. In Sect.~\ref{sec:2}, we provide an abridged summary of model comparison techniques.  The summary of DAMA results in their latest data release paper can be found in Sect.~\ref{sec:3}. Our analysis and results are described in Sect.~\ref{sec:4}. \rthis{Our results of searches for higher harmonics are described in Sect.~\ref{sec:higherharmonics}. We also carry out sensitivity analysis of a dark matter experiment with similar backgrounds as ANAIS-112 or COSINE-100 in confirming the DAMA annual modulation signal in Sect.~\ref{sec:sensanalysis}.} We conclude in Sect.~\ref{sec:conclusions}. Our analysis codes have been made publicly available and the {\tt github} link for the same is provided in Sect.~\ref{sec:conclusions}.

\section{Introduction to Model Comparison Techniques}
\label{sec:2}

 In recent years a large number of both Bayesian, frequentist, and information theory based  model comparison  techniques  (originally developed by the statistics community) have been applied to a variety of problems in astrophysics and cosmology involving hypothesis testing or comparing which of two models is favored and quantifying its statistical significance. 
 For our purpose, we shall employ all the available techniques at our disposal to address the significance of the DAMA annual modulation claim. We briefly recap these techniques below. More details on each of these tests can be found in various reviews~\cite{Lyons,Liddle,Liddle07,Weller},
 and proof of principles applications of some of these techniques to problems in astrophysics and cosmology can be found in Refs.~\citep{Shi,Shafer,Heavens,Kulkarni,Liu,Ganguly,Keitel}.

\begin{itemize}

\item {\bf Frequentist Test}: This method for model comparison is sometimes also known as  the Likelihood ratio test~\citep{Weller}. The first step in this, involves parameter estimation for a given hypothesis, usually by minimizing the $\chi^2$ functional between the data and model. Then, based on   the best-fit $\chi^2$ between a given model and data and from the  degrees of freedom, one calculates the goodness of fit for each model, given by the $\chi^2$ probability distribution function~\citep{astroml}. This allows one to penalize models with complexity.

The best-fit model between the two is the one with the larger value of $\chi^2$ goodness of fit. If the two models are nested, then from Wilk's theorem~\cite{Wilks}, the difference in $\chi^2$  between the two models satisfies a $\chi^2$ distribution with degrees of freedom equal to  the difference in the number of free parameters for the two hypotheses~\citep{Lyons,Weller}. Therefore, the $\chi^2$ c.d.f. can be used to quantify the $p$-value of a given hypothesis.
From the $p$-value, one usually calculates a $Z$-score or number of sigmas, which is the number of standard deviations that a Gaussian variable would fluctuate in one direction to give the corresponding $p$-value~\cite{Cowan11,Ganguly}. 
This  test has been used for a variety of applications in  astrophysics and cosmology~\cite{Shafer,Desai16b,Kulkarni,Ganguly}. We note that the DAMA collaboration also uses this test in their data release papers to assess the significance of their annual modulation claim.

\item {\bf Akaike Information Criterion}:

 The Akaike Information Criterion (AIC) is used to penalize for any  free parameters  to avoid overfitting. AIC is an approximate  minimization of Kullback–-Leibler information entropy, which estimates the distance between two probability distributions~\citep{Liddle07} and is given by:
\begin{equation}
\rm{AIC} = -2\ln{}{\mathcal{L}_{max}} + 2p,
\label{eq:AIC}
\end{equation}
where  $p$ is the number of free parameters, and $\mathcal{L}$ is the likelihood. A preferred model in this test is the one with the smaller value of AIC between the two hypothesis. There is no formal method to evaluate the $p$-value from the difference in AIC between the two models~\footnote{See however Ref.~\cite{Shafer}, which posits a significance based on $\exp(-\Delta AIC/2)$}, and usually qualitative strength of evidence rules are used~\citep{Shi} to estimate the relative significance of the favored model.

\item {\bf Bayesian Information Criterion}:
The Bayesian Inference Criterion (BIC) is also used for penalizing the use of extra parameters and is an approximation to the Bayes factor. It is given by~\cite{Liddle}:
\begin{equation}
\rm{BIC} = -2\ln{}{\mathcal{L}_{max}} + p \ln N, 
\label{eq:BIC}
\end{equation}
where all the parameters have the same interpretation as in Eq.~\ref{eq:AIC}.
Similar to AIC, the model with the smaller value of BIC is the preferred model. 
The significance is estimated qualitatively in the same way as for AIC.  


Besides AIC and BIC, other information criterion based tests such as Takeuchi information criterion and Deviance information criterion~\citep{Liddle07} have also been proposed. But their computation as well as interpretation is not straightforward, and hence we do not implement them in this work.

\item {\bf Bayesian Model Selection}:
Apart from the information theory-based techniques such as AIC and BIC, Bayesian model comparison techniques~\cite{Trotta} have also been extensively used for model comparison in astrophysics and cosmology~\cite{Shafer,Heavens,Trotta,Pitkin}. The key quantity,  which needs to be computed to compare two models ($M_1$ and $M_2$) is the Bayesian odds ratio, given by:
\begin{equation}
O_{21} = \frac{P(M_2 | D)}{P(M_1|D)},
\end{equation}
\noindent where $P(M_2 | D)$ is the posterior probability for $M_2$  given data $D$, and similarly for  $P(M_1 | D)$. 
The posterior probability for a general model $M$ is given by
\begin{equation}
P(M |D) = \frac{P(D| M) P (M)}{P (D)},
\label{eq:bi}
\end{equation}
where $P(M)$ is the prior probability for the model $M$, $P(D)$ is the probability for the data $D$ and $P(D|M)$  is the marginal likelihood or Bayesian evidence for model $M$ and it quantifies the probability that the data $D$ would be observed if the model $M$ were the correct model. $P(D|M)$  is given by:
\begin{equation}
P(D |M) = \int P (D | M,\theta) P(\theta | M) d\theta, 
\end{equation}
where $\theta$ is a vector of parameters,  which parametrizes the model $M$. If the prior probabilities of the two models are equal, the odds ratio can be written as:
\begin{equation}
B_{21} = \frac{\int P(D | M_2, \theta_2)  P(\theta_2)d \theta_2}{\int P (D | M_1, \theta_1) P (\theta_1) d\theta_1}, 
\label{eq:bayesfactor}
\end{equation}
The quantity $B_{21}$ in Eq.~\ref{eq:bayesfactor} is known as Bayes factor. We shall  compute this quantity in order to obtain a Bayesian estimate of the statistical significance.

Unlike the frequentist model-comparison technique, there is no quantitative way to rank between two models. However, similar to AIC and BIC, a qualitative criterion based on Jeffreys scale is used to interpret the odds ratio or Bayes factor~\cite{astroml,Trotta}. A value of $> 10$ represents strong evidence in favor of $M_2$,  and a value of $> 100$ represents decisive evidence~\cite{astroml,Trotta}. 
\end{itemize}

\section{Summary of DAMA/LIBRA phase2 results}
\label{sec:3}

Here, we provide a succinct summary of  the key results in the latest data release paper from DAMA/LIBRA~\citep{Dama18}, which we later try to reproduce. For more details, the reader can consult Ref.~\citep{Dama18}.

The total data presented in Ref.~\citep{Dama18} for 1-3 keV   and 1-6 keV corresponds to six annual cycles of the DAMA/LIBRA phase II, with a total exposure of 1.13 ton year. The data in the 2-6 keV energy bin corresponds to the DAMA/LIBRA phase II data combined with data from DAMA/NaI and DAMA/LIBRA phase I (collected over 14 annual cycles), with a total exposure of 2.46 ton years.
We note that there is an overlap in energy between the different intervals.
In previous DAMA papers~\citep{DAMA03,Dama08,DAMA10,Dama13} containing a combination of   data from phase I of DAMA/LIBRA experiment as well as DAMA/NaI, single-hit rates were also presented in the 2-4 keV, 2-5 keV, and 2-6 keV
energy intervals. 

The DAMA Collaboration looks for an annual modulation signature indicative of a dark matter signal, which must satisfy some basic characteristics such as a period of one year, a phase peaking roughly around  
June 2nd, occurring only in single-hit events, and the modulation amplitude having at most 7\% of the DC amplitude~\citep{Dama18}. The dataset used for this search consists of residual rates of the single-hit scintillation events in each energy interval, after subtracting a constant background. This constant background is obtained after averaging the rate per detector in each energy interval over all the annual cycles~\cite{DAMA04}. The dataset has been binned into unequal size binned intervals. \rthis{The median bin size is 19.9, 20.43, 25.0,25.27, and 21.78 for the 2018 1-6 keV, 2018 1-3 keV, 2012 2-5 keV, 2012 2-4 keV and  2-6 keV data (from 2012 and 2018) respectively.}
The single-hit residual rates in DAMA/LIBRA phase 2 have been fit to the following cosine function 
\begin{equation}
H_1(x) = A\cos\omega(x-t_0),
\label{eq:cosine}
\end{equation}
where $\omega=\frac{2\pi}{T}$ corresponding to period $T$.
Their best-fit values are given by $T \sim 1$ year,  $t_0 \sim 152.5$ days (at around  June 2), and $A\sim 0.01$ cpd/kg/keV. The reduced $\chi^2$
for all the fits are close to 1 and can be found in Table 1 of Ref.~\citep{Dama18}. The statistical significance of this annual modulation corresponds to a $Z$-score of 8.0$\sigma$, 9.6$\sigma$, and  8.7$\sigma$ in 1-3 keV, 1-6 keV, and 2-6 keV energy intervals respectively. When the 2-6 keV data is combined with data from DAMA/NaI and phase 1 of DAMA/LIBRA, the significance gets enhanced to 12.9$\sigma$. In the 2012 data release from DAMA/LIBRA, the significance of annual modulation in 2-4 and 2-5 keV intervals was $7.6\sigma$.


\section{Analysis and Results}
\label{sec:4}
For our analysis, the data and the associated errors  in  the plots (from Ref.~\citep{Dama18}) in each energy interval have been digitized~\footnote{The raw DAMA data presented in their papers is not publicly available to the best of our knowledge}. The published data consists of single-hit residual rates  in three different energy intervals. Each data point consists of experimental errors  in the residual rates, as well as the associated time widths as errors in the independent variable. The abscissa values range from 6200 to 8300 days for 1-3 and 1-6 keV energy intervals and from 3000 to 8300 days for 2-6 keV interval(cf. Figs.~2 and 3 in Ref.~\citep{Dama18}).

In order to independently evaluate the significance of the apparent annual modulation present in the signal, we use  both a constant function, given by
\begin{equation}
H_0(x) = k,
\label{eq:const}
\end{equation}
and the cosine model  in Eqn.~\ref{eq:cosine} as two different models, and compare the best fits for both of them.
Initially, we estimate the best-fit parameters  for both these hypotheses. This is the first step towards model comparison. The constant function is then considered as the null hypothesis and model comparison is carried out accordingly.

\subsection{Parameter Estimation}

The first step in model comparison involves estimating the best-fit parameters of a given model. For this purpose, we construct a $\chi^2$ function between the data and a given model. We also include the errors in the time bin. Although, no details of the $\chi^2$ functional used for minimization is provided in Ref.~\citep{Dama18} or in earlier papers, most likely their $\chi^2$ does not include the errors in the abscissa. To incorporate the error in the independent time variable, we follow the method of Ref.~\citep{Weiner} and the total error ($\sigma_{t}$) is given by:
\begin{equation}
\sigma_t= \sqrt{\sigma_H^2+\sigma_x^2 \left(\frac{\partial H }{\partial x}\right)^2},
\label{eq:toterror}
\end{equation}
\noindent where $\sigma_H$ is the error in residual rate (obtained by digitization of the relevant plots in Ref.~\citep{Dama18}), $\sigma_x$ is the error in time variable and 
$\frac{\partial H }{\partial x}$ is obtained from Eq.~\ref{eq:cosine}. We assume that the error bars provided by DAMA/LIBRA collaboration are robust and all systematic terms have been included, and we don't need to fit for unknown systematic error terms (for eg. Ref~\citep{Desai16b} in the context of periodicity in measurements of Newton's constant). We also assume that the amplitude $A$ in Eqn.~\ref{eq:cosine} does not vary with time. Time-dependence of $A$ is explored in Ref.~\citep{Freese17}. The $\chi^2$ functional is then given by:
\begin{equation}
\chi^2= \sum_{i=1}^N\left( \frac{y_i-H(x)}{\sigma_t}\right)^2
\label{eq:chisq}
\end{equation}
\noindent where $H(x)$ is defined in Eq.~\ref{eq:cosine} and $y_i$ denote the residual single-hit DAMA rates in each time bin $i$. For finding the minimum value of $\chi^2$, we have kept free all the three parameters of the cosine function.

\begin{figure*}
\includegraphics[width = .75\textwidth]{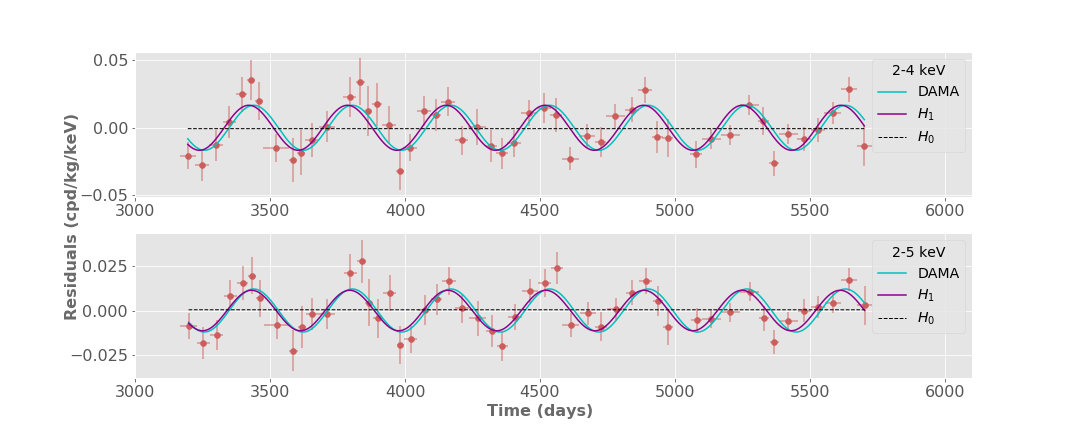}
\includegraphics[width = .75\textwidth]{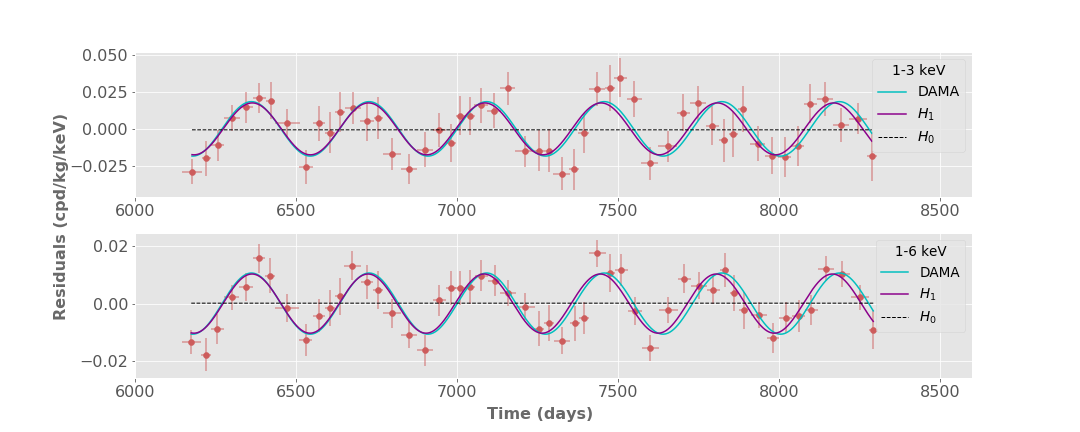}
\includegraphics[width = \textwidth]{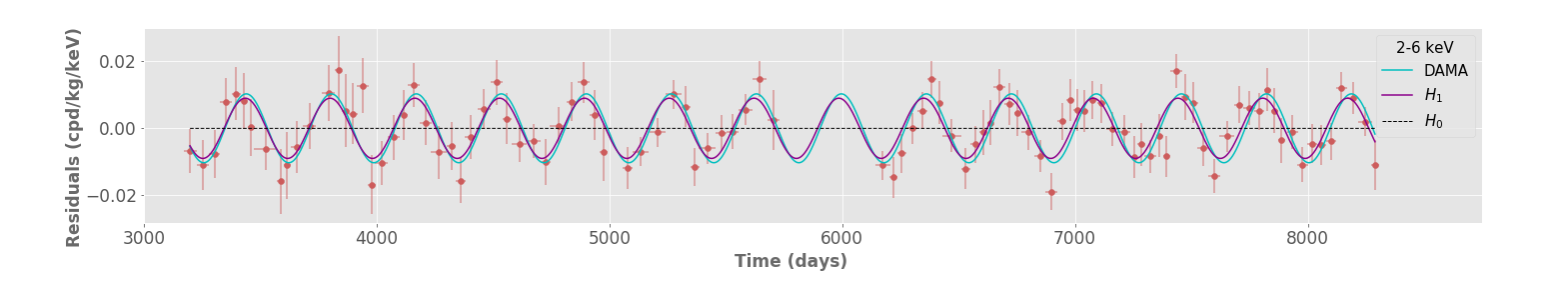}

\caption{The DAMA data points (red) in each energy range, namely 2-4 keV, 2-5 keV, 1-3 keV, 1-6 keV,  and 2-6 keV, are overlaid with both the corresponding calculated fits of $H_1(x)$ (Eq.~\ref{eq:cosine}) to the data, as well as the fits the DAMA/LIBRA collaboration calculated, as depicted in the first set of values in Table 1 of Ref.~\cite{Dama18}. The sinusoidal fit calculated by this work is depicted in purple while that calculated by the DAMA/LIBRA collaboration is shown in cyan. The calculated fit for $H_0(x)$ is also shown.}
\label{fig:boat2}
\end{figure*}


For each of the given sets of data in the 2-4, 2-5, 1-3, 1-6, 2-6 keV  range, the data is fit to both the hypotheses by minimizing the $\chi^2$ in Eq.~\ref{eq:chisq}  to obtain the best-fit parameters. The parameters obtained for all the  energy ranges are displayed in  Table~\ref{table:2}
for the cosine and also the constant models. The values of parameters obtained for the cosine model, $H_{1}(x)$ agree within $1\sigma$ compared to those found by  the DAMA collaboration~\cite{Dama18}. Fig.~\ref{fig:boat2} represents the best-fit constant value as obtained for $H_0(x)$ for each energy range, along with the best fit cosine wave obtained, in comparison to the cosine wave with the best-fit parameters presented by the DAMA collaboration \cite{Dama13}. 

\subsection{Frequentist Model Comparison}
From the figures, we see that the cosine model appears to fit the data very well for all the five sets of data. To quantify this goodness of fit, we calculate the value of $\chi^2$/DOF,  and additionally the value of $\chi^2$ probability as defined in Section \ref{sec:2}, hereafter denoted as GOF for goodness of fit. It is expected that for reasonably good fits, the value of $\chi^2$/DOF will be close to 1. The values of $\chi^2$/DOF and GOF are indicated in Table~\ref{table:4}. We see that the values of $\chi^2$/DOF are much closer to 1 for the cosine model than for the constant model. Additionally, the GOF values for the cosine model are of the order of $10^{-2}$, whereas for the constant model the values are $\mathcal{O}(10^{-6})$. Thus, we can reasonably conclude that the cosine model provides a better fit to the data. Further, the time period of the observed fit is very close to 1 year, and the phase offset, $t_0$ is comparable to the required 152.5 days. Thus, the frequentist test strongly supports annual modulation.

\begin{table*}
\begin{center}
\begin{tabular}[t]{|c|c|c|c|c|}
\hline
\multirow{2}{*}{\textbf{Energy}} & \multicolumn{3}{c}{$\mathbf{H_1}$} & \multicolumn{1}{|c|}{$\mathbf{H_0}$}\\ 

 \textbf{Interval}&$\mathbf{A}$ (cpd/kg/keV) & $\boldsymbol\omega$ (radians/day) & $\mathbf{t_{0}}$ (days) & $\mathbf{k}$ (cpd/kg/keV)\\ [0.5ex]
 \hline\hline
 \multicolumn{1}{|c|}{2012 data}&&&& \\
 2-4 keV & 0.0167 & 0.0172 & 131.69 & $-6.42\times10^{-4}$  \\
 2-5 keV & 0.0115 & 0.0173 & 161.84 & $4.20\times10^{-4}$\\
\hline
 \multicolumn{1}{|c|}{2018 data}&&&& \\
 1-3 keV& 0.0175 & 0.0174 & 224.17 & $-6.41\times10^{-4}$ \\
 1-6 keV& 0.0102  & 0.0174  & 234.13  & $1.50\times10^{-4}$ \\
\hline
\multicolumn{1}{|c|}{2012 + 2018 data}&&&& \\
 2-6 keV&	0.0090  & 0.0172 &156.67  & $-2.16\times10^{-5}$\\
\hline
\end{tabular}
\end{center}
\caption{Best-fit values for different parameters of the cosine model (cf. Eqn.~\ref{eq:cosine}) as well as the constant model (cf. Eqn.~\ref{eq:const})  to  the DAMA data points in different energy intervals.}
\label{table:2}
\end{table*}

As noted in Sect. \ref{sec:2}, it is possible to take advantage of the fact that the constant model is nested within the cosine model (when $A=0$), and apply Wilk's theorem to evaluate the statistical significance of the cosine model in comparison to the null hypothesis. Thus, the difference in $\chi^2$ between the constant and the cosine models follows a $\chi^2$ distribution with DOF = 2~\cite{Lyons,Weller}. From the distributions obtained, we calculate the $p$-value of the cosine model for each set of data separately, from the cumulative distribution of  the $\chi^2$ functional. The $p$-values obtained are depicted alongside the difference in $\chi^{2}$ values in Table~\ref{table:5}. The $p$-value can be interpreted as the probability that we would see data that favours the cosine model by chance, given that the null hypothesis is true. Thus, the smaller values of the $p$-value indicate that the cosine model should be favored over the constant model. For each of the energy ranges, the corresponding significance (or $Z-$ score) is calculated~\citep{Cowan11,Ganguly}.
Its value is given alongside the $p$-value. These can be found in Table~\ref{table:5}. We note that the significance values are comparable (within $\pm 1\sigma$) to the values indicated in the DAMA/LIBRA findings, presented in Refs.~\cite{Dama13,Dama18}.


\begin{table*}
\begin{center}
\begin{tabular}{|c|c|c|c|c|c|c|c|c|c|c|}
    \hline
    {} & \multicolumn{4}{c}{{2012 data}} & \multicolumn{4}{|c|}{{2018 data}}&
    \multicolumn{2}{c|}{2012 + 2018 data}\\
    \cline{2-11}
    \textbf{}  &
    \multicolumn{2}{c|}{{2-4 keV}}&
    \multicolumn{2}{c|}{{2-5 keV}}&
    \multicolumn{2}{c|}{{1-3 keV}} &
    \multicolumn{2}{c|}{{1-6 keV}}&
    \multicolumn{2}{c|}{{2-6 keV}}\\
    
    {\textbf{Model}}&{Cosine}&{Constant}&{Cosine}&{Constant}&{Cosine}&{Constant}&{Cosine}&{Constant}&{Cosine}&{Constant}\\
    
    \cline{2-11}

    $\mathbf{\chi^2}$\textbf{/DOF} & 35.5/47 & 110.5/49 & 32.4/47 & 99.8/49 & 43.7/49 & 110.4/51 & 38.0/49 & 153.6/51  & 59.2/99  &  198.3/101   \\ 
    \textbf{GOF}  & 0.023 & $3.55\times10^{-7}$ & 0.014 & $6.85\times 10^{-6}$ & 0.039 & $8.08\times10^{-7}$ & 0.025 & $1.13 \times 10^{-12}$    & $1.82\times10^{-4}$  & $6.57\times10^{-9}$ \\

    \hline
\end{tabular}
\end{center}
\caption{Summary of best-fit $\chi^2$ per degrees of freedom, GOF (obtained using the $\chi^2$ p.d.f for the cosine and constant model) for the different energy intervals.  The corresponding values for these by the DAMA/LIBRA collaboration can be found in Ref.~\cite{Dama18} for the 2018 data and in Tables 2 and 3 for the 2012 data in Ref.~\cite{Dama13}. }
\label{table:4}
\end{table*}

\subsection{Information Criteria}
Subsequently, we proceed to calculate the difference in AIC and BIC values between the null hypothesis and the cosine model, where the values are calculated using Eqns.~\ref{eq:AIC} and \ref{eq:BIC}. In general, the model with the smaller AIC and BIC values is preferred. The $\Delta$AIC and $\Delta$BIC  values can be found in Table~\ref{table:5}, where the cosine model has the smaller value. From  the difference in AIC and BIC values, we  evaluate the significance using  the qualitative ``strength of evidence rules'' provided in Ref.~\cite{Shi}. As all the values of $\Delta$AIC and $\Delta$BIC are in the range of 50-100, they provide \textit{very strong evidence} against the null hypothesis,  according to the aforementioned scale. 

\begin{table*}
\begin{center}
\begin{tabular}{|c|c|c|c|c|c|c|}
    \hline
    \textbf{Energy interval}  &
    $\mathbf{\Delta\chi^2}$ &
    \textbf{$p$-value}&
    \textbf{$Z$-score}&
    \textbf{$\boldsymbol\Delta$AIC}&
    \textbf{$\boldsymbol\Delta$BIC}&
    \textbf{Bayes Factor}\\ \cline{2-5}

    \hline\hline
    \multicolumn{1}{|l|}{2012 data} & & & & & & \\
    2-4 keV & 75.0 & $5.18\times10^{-17}$ & $8.3\sigma$ & 59.68 & 55.85 & $4.30\times10^{10}$\\
    2-5 keV & 67.3 & $2.37\times10^{-15}$ & $7.8\sigma$ & 53.23 & 49.41 & $3.29\times10^8$\\
    \hline
    \multicolumn{1}{|l|}{2018 data}&&&&&& \\
    1-3 keV & 66.8 & $3.16\times10^{-15}$ & $7.8\sigma$ & 55.26 & 51.35 & $9.46\times10^9$ \\
    1-6 keV& 115.5 & $8.17\times10^{-26}$ &  $10.4\sigma$ & 99.46 & 95.56 & $1.26\times10^{18}$\\
    \hline
    \multicolumn{1}{|l|}{2012 + 2018 data}&&&&&& \\
    2-6 keV & 139.2 & $6.02\times10^{-31}$ & $11.5\sigma$ & 117.17 & 111.92 & $2.82\times10^{17}$\\
    \hline
\end{tabular}
\end{center}
\caption{Summary of model comparison results for the sinusoidal variation to the DAMA data compared to the constant fit as the null hypothesis, using all three model comparison techniques. The first two  columns depict the $\Delta \chi^2$ value between the constant fit~\ref{eq:const} and the cosine model~\ref{eq:cosine} and the $p$-value obtained using Wilk's theorem. The third column indicates the $Z$-score. The corresponding $Z$-scores found by the DAMA collaboration can be found in Table 2 of Ref.~\cite{Dama18} for the 2018 data and in Table 4 of Ref.~\cite{Dama13}. The next two columns indicate the difference in AIC and BIC between the two models. Finally, the last  column shows the Bayes factor between the two models. As we can see, the sinusoidal model is decisively favored over the constant model using all the three techniques.}
\label{table:5}
\end{table*}

\subsection{Bayesian Analysis}

Finally, we proceed to calculate the Bayesian odds ratio $B_{21}$ for the $M_2$ model in comparison to the $M_1$ hypothesis. For this purpose, we consider the null hypothesis to be $M_1$ and the cosine model to be $M_2$. Each  set of points in a given  energy range is separately considered as data $D$, and $B_{21}$ is calculated for each energy range, namely 2-4 keV, 2-5 keV, 1-3 keV, 1-6 keV, and 2-6 keV.

We initially  need to calculate the Bayesian evidence for each of the two models.
The first step in the calculation of Bayesian evidence is the likelihood of the data, given the model and a set of parameters, and for this purpose we use a Gaussian likelihood
\begin{equation}
P(D|M,\theta)= \prod_{i=1}^N \frac{1}{\sigma_t\sqrt{2\pi}} \exp\left[-\frac{1}{2}\left(\frac  {y_i-H(x)}{\sigma_t}\right)^2\right], 
\end{equation}
\noindent where $\sigma_t$ is the total error defined in Eq.~\ref{eq:toterror}, $y_i$ denote the DAMA data and $H(x)$ represents the model, which in this case is either the constant or sinusoidal model.

The Bayesian priors chosen are the constant $k$ and the amplitude $A$ uniform over [0,$A_{max}$] (where $A_{max}$ is the maximum absolute value of the residual rate in the particular data set in consideration, with the mean of the $A_{max}$ values from the five data sets being 0.0270 cpd/kg/keV), $\omega$ uniform over [0,2$\pi$/365.25] and the phase $\omega t_0$ uniform over [0,2$\pi$]. 
-To calculate the Bayesian evidence for both the hypotheses, we used the {\tt Nestle} package in {\tt Python}~\footnote{See http://github.com/kbarbary/nestle}, which  uses the nested sampling algorithm~\cite{multinest,Mukherjee}.

The Bayes factor is given by the ratio of Bayesian evidence for the cosine and constant model hypothesis. The Bayes factors for different combinations of datasets and energy intervals are summarized in Table~\ref{table:5}.
We interpret the Bayes factor using the Jeffrey's Scale~\cite{Trotta}. As the values of Bayes factor are well above 100 for all data sets, we conclude that they provide \textit{decisive evidence} against the null hypothesis.

\section{Searches for higher harmonics}
\label{sec:higherharmonics}
\rthis{We now address the question as to  whether higher harmonics are present in DAMA/LIBRA data, in addition to the cosine signal with a period of one year. This question has been previously addressed in some of the papers by the  DAMA collaboration~\cite{DAMA10,Dama13,Bernabei13a,Dama18} and also by other authors~\cite{Sturrock}. The DAMA collaboration has used an ordinary periodogram  as well as  the Lomb-Scargle (L-S, hereafter) periodogram  to look for higher harmonics. The L-S
periodogram~\cite{Lomb,Scargle} is an extension of the ordinary periodogram, and is a widely used technique
in astronomy and particle physics to look for periodicities in unevenly sampled datasets. A recent comprehensive review of the L-S periodogram (including all its variants) can be found in Ref.~\cite{vanderplas}.}

\rthis{The DAMA collaboration could not find any additional peaks (other than one year) in these periodograms for the data in the 2-6 keV energy intervals, when they scanned the frequencies  up to the Nyquist limit of  0.5/per day (corresponding to a period of 2 days). They also found that the 6-14 keV energy range does not contain any peaks in the periodogram, and  is consistent with only noise~\cite{Dama18,Bernabei13a}.}

\rthis{An independent L-S analysis of the digitized DAMA data was carried out in Ref.~\cite{Sturrock}, wherein the authors also criticized some aspects of the analysis procedure of the DAMA collaboration~\cite{DAMA10}. Sturrock et al have pointed out that the DAMA L-S analysis does not incorporate the experimental errors. Sturrock et al~\cite{Sturrock}  independently analyzed the DAMA data until 2010 using three different variants of the standard L-S periodogram  and found significance for annual modulation at the level of 5.8$\sigma$ to 6.6$\sigma$. They also pointed out some limitations in searching for higher harmonics in the publicly available DAMA data, which is binned. One caveat is that the maximum frequency which can be searched for is determined by the size of the time bin.  Another problem in searching for higher harmonics in the DAMA data is that 
the residuals are calculated by subtracting the mean on an annual basis, but the mean value has not been made publicly available. Nevertheless, their analysis also did not find any evidence for  any higher harmonics~\cite{Sturrock}.}

\rthis{We did an independent search for higher harmonics in the 1-6 keV and 2-6 keV energy range. For this purpose, we used the generalized~\cite{Kurster} or floating-mean~\cite{Cumming}  periodogram, in which an offset term is added at each frequency. This generalized  periodogram has been shown to be more sensitive than the normal L-S periodogram~\cite{vanderplas}. We have previously used this version of the generalized L-S periodogram  to look for periodicities in beta decay rates and solar neutrino fluxes~\cite{Liu,Tejas}. For our analysis, we implemented this generalized L-S  periodogram using the {\tt LombScargle} class in the  {\tt astropy}~\cite{astropy} module. The DAMA collaboration and Ref.~\cite{Sturrock} have looked for higher harmonics from the periodograms of the raw DAMA residual data itself, which is already indicative of annual modulation. In principle, this data can also be  used to detect additional periodic components with high amplitudes. However, since the L-S periodogram is essentially equivalent to a single-term sinusoidal fit to the data at each frequency, the generalization  to multiple frequencies can lead to spurious periodogram peaks~\cite{astroml,vanderplas}
Therefore, as a complementary approach to the previous analysis in Ref.~\cite{Sturrock} and DAMA collaboration~\cite{Dama18}, we search for higher harmonics from the residuals, obtained from the subtracting the best fit cosine with a period of one year. We however note that we also applied the generalized L-S periodogram on the raw DAMA data, but we do not report the results here, as they are not different than the analysis done on the residuals, which we now present.}

\rthis{These residuals are calculated  by subtracting the best-fit value for the cosine model from the DAMA data, presented in Table~\ref{table:2}. Since the DAMA data is binned, the maximum frequency which we can search for is limited by the bin size. The median time bin width for the DAMA data in the 1-6 keV and 2-6 keV interval is equal to 19.9 and 21.8 days respectively. Since both these values  are close to 20 days, the maximum frequency which we search in both the energy bins is equal to a period of 20 days, or a frequency of  18 /per year. If the DAMA collaboration released their data in one-day bins, one could have searched for periodicities at  higher frequencies as well, similar to what was done in some of the papers by the DAMA collaboration.}

\begin{figure*}
\includegraphics[width = 0.5\textwidth]{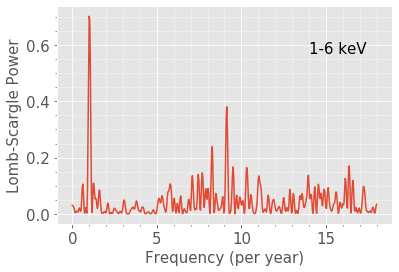}
\includegraphics[width =0.5\textwidth]{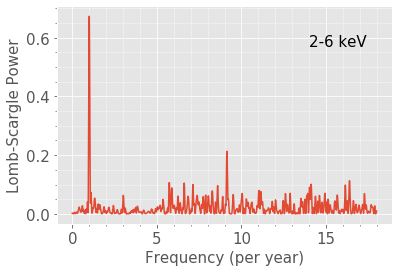}
\caption{Generalized Lomb-Scargle periodogram~\cite{Kurster} of the DAMA best-fit residuals in 1-6 keV (top figure) and 2-6 keV (bottom figure). These residuals have been computed by subtracting the best-fit values for the period of 1 year. The residuals still show the peak at 1 year. In addition to this, both the periodograms show an additional peak at a frequency of  9.13 per year (period of 40 days), with $p$-values of 0.004 (2.64$\sigma$) and 0.0066 (2.47$\sigma$) in 1-6 and 2-6 keV energy intervals. Therefore the significance is marginal and furthermore, the period of 40 days is close to the median DAMA bin size of 20 days.}
\label{fig:LS}
\end{figure*}

\rthis{Our results are shown in Fig.~\ref{fig:LS} for both 1-6 keV and 2-6 keV energy intervals. The normalization convention, which we have used for L-S periodogram can be found in 
Ref.~\cite{vanderplas}. For both these energy intervals, we can again see a peak at a frequency of 1 year. The only other peak which we see in the  1-6 keV interval is at a frequency of 9.13 per year corresponding to the L-S power of 0.38. This corresponds to a $p$-value  of 0.004 with a $Z$-score of 2.64$\sigma$. The $p-$value has been calculated using the prescription of Baluev~\cite{Baluev}. This same peak at 9.13 per year is also seen in the 2-6 keV energy region with a L-S power of 0.21. The $p$-value for this peak is 0.0066 with a $Z$-score of $2.47\sigma$. We note however that  this frequency of 9.13/year corresponds to a period of about 40 days and is close to the  median bin width of the DAMA data. Furthermore, it is not very significant. More data in smaller time widths is needed to get a more robust estimate of its significance and to see if this peak persists with additional data.}

\section{Sensitivity of COSINE-100 and ANAIS-112 searches}
\label{sec:sensanalysis}
\rthis{Two other dark matter experiments sensitive to annual modulation, which use NaI as target and  designed to test the DAMA results, namely COSINE-100 and ANAIS-112 have started taking data and released preliminary results with a lifetime of about a  year ~\cite{Cosine,Anais}. We have also carried out an independent search for annual modulation with the same data using all the four model independent techniques used here~\cite{Krishak1,Krishak2}. Currently, the ANAIS-112 and COSINE-100 exposures are not very large to robustly discriminate between a background only versus signal only hypothesis. However, this can change with more accumulated exposure.
Therefore, we investigate using numerical experiments on how the significance  would change as a function of exposure time, if the data contained a real sinusoidal signal as well as if the data contained only background. We note that more detailed sensitivity analysis of  the DAMA parameter space has been done by the COSINE-100~\cite{Cosinesens} and  ANAIS-112~\cite{Anaissens,Anaissens2} collaborations, which take into account experimental details such as detection efficiency, quenching factors. etc. Here, we  investigate  using synthetic data how the metrics we have used for model comparison with real data (frequentist significance, AIC, BIC, Bayes factor) would help adjudicate between a true signal and noise, as  a function of exposure time.}

\rthis{Since both COSINE-100 and ANAIS-112 experiments have similar detector fiducial mass ($\mathcal{O}$ (100 kg)) and  background rates (2-4 cpd/kg/keV), we consider the sensitivity of an hypothetical dark matter experiment designed  to detect annual modulation, which  would mimic the behaviour of  both these experiments.  Our numerical experiments should give an  idea of how the model comparison metrics we have considered in this work change as a function of the exposure time and are sensitive to a real signal. We first describe the results, assuming  the detector data stream only contains noise. We injected a cosine signal with the same period  and phase as detected by DAMA collaboration and with different amplitudes (to emulate different signal to noise ratios) in order to ascertain the sensitivity as a function of signal to background ratio. 
We now present results for model comparison analysis for the same.}

\subsection{Results with only background}
\rthis{We first describe the recipe  used  to generate five years worth of simulated detector background, which can be used to look for dark-matter induced signals.
Since the search for dark matter-induced signals is done using the residual data, which is obtained by subtracting the exponential background, we create a synthetic dataset of the residual data, based on  the observed ANAIS-112 residual data, which is publicly available~\cite{Anais}. Since the COSINE-100 background rate (2.7 cpd/kg/keV/ is of the same order of magnitude as the ANAIS-112 background rate (2-4 cpd/kg/keV), this synthetic background only dataset should also be representative of the COSINE-100 experiment.
We assume that the data has been binned in 10-day intervals (similar to ANAIS-112). The time binning assumed for the COSINE-100 data is also similar  (15 days). One simplification, we have made is that our simulated  experiment consists of only one module. In reality, both COSINE-100 and ANAIS-112 consist of multiple modules, and data from all these modules is used concurrently to search for a cosine signal.}

\rthis{For five years worth of synthetic data, we generated 184  normal random deviates  with mean of zero and scale parameter same as the COSINE-100 residual background, equal to 0.035 cpd/kg/keV (obtained from  Fig.~2 of Ref.~\cite{Anais}). Since the distribution of the ANAIS-112 residuals is close to a Gaussian distribution, therefore our simulated dataset should emulate the observed backgrounds in these experiments.
Each simulated data point encapsulates the background rate within a 10-day interval. To simulate the  error distribution of the background data, we generated normal deviates with mean obtained from the median error of the ANAIS-112 residual rates (equal to 0.031 cpd/kg/keV) and scale parameter, which mimics the distribution   of the errors in the same data.
The above simulated five years  data-stream  was then used for model comparison. We used a subset of the above data to do a similar analysis for one and three years worth of data. }

\rthis{We now fit this data to both a  cosine signal (with the period fixed to the DAMA best fit value of a year and both amplitude and phase as free parameters) as well as a constant background rate and carry out model comparison in the same way as done for the real data. We note that the best-fit maximum likelihood estimate for a constant background is given by the weighted mean~\cite{Bevington} of the simulated data along with the simulated error distribution. Our results are shown in Table~\ref{table:simnoise}. We find that although all the model comparison tests prefer the constant model over a cosine model, the significance for both the models is comparable for all the three exposures  considered, and none of the tests can robustly discriminate between the two hypotheses. One possible reason for this is that the fluctuations of  the simulated background is large and a constant background does not adequately fit all the data points. For a smaller background scale parameter and with smaller error bars, the metrics would more robustly favor a constant model.}

\begin{table*}
\begin{center}
\begin{tabular}{|c|c|c|c|c|c|c|}
    \hline
    {} & \multicolumn{2}{c}{{Year 1}} & \multicolumn{2}{|c|}{{Year 3 }}&
    \multicolumn{2}{c|}{Year 5}\\
    \hline
    {\textbf{Model}}&{Cosine}&{Constant}&{Cosine}&{Constant}&{Cosine}&{Constant}\\
    
    \hline

    $\mathbf{\chi^2}$\textbf{/DOF} & 59/36 & 57/37 & 155/108 & 154.1/109 & 237/181 & 236/182    \\ 
    \textbf{GOF}  & 0.0022 & 0.0043 & 0.0003 & 0.0004 & 0.00004 & 0.0005  \\
    \cline{2-7}
    \textbf{$\Delta \chi^2$ ($p$-value)} & \multicolumn{2}{c|}{1.98(0.15)} & \multicolumn{2}{c|}{0.8 (0.36) } & \multicolumn{2}{c|}{0.9 (0.34)} \\
    
\textbf{$Z$-score} & \multicolumn{2}{c|}{1.0$\sigma$} & \multicolumn{2}{c|}{0.34$\sigma$} & \multicolumn{2}{c|}{0.4$\sigma$} \\
    \hline
    \textbf{AIC(const) - AIC(cos)} & \multicolumn{2}{c|}{-2.98} & \multicolumn{2}{c|}{-1.8} & \multicolumn{2}{c|}{-1.9} \\
    \hline
    \textbf{BIC(const) - BIC(cos)} & \multicolumn{2}{c|}{-5.5} & \multicolumn{2}{c|}{-5.5} & \multicolumn{2}{c|}{-6.1} \\
    \hline
     \textbf{Bayes factor} & \multicolumn{2}{c|}{0.24} & \multicolumn{2}{c|}{0.87} & \multicolumn{2}{c|}{0.76} \\
     \hline
\end{tabular}
\end{center}
\caption{\rthis{Summary of model comparison tests on simulated data of a hypothetical dark matter experiment (similar to COSINE-100 or ANAIS-112 designed to detect annual modulation) consisting of only background. The backgrounds have been obtained by simulating a normal distribution with mean of zero and  scale parameter  same as that of the ANAIS-112 residual rates~\cite{Anais} (equal to 0.035 cpd/kg/keV). We conclude that although all the model comparison tests prefer a constant model, the significance is not strong enough to decisively favor any one model over the other.}}
\label{table:simnoise}
\end{table*}

\begin{table*}
\begin{center}
\begin{tabular}{|c|c|c|c|c|c|c|}
    \hline
    {} & \multicolumn{2}{c}{{Year 1}} & \multicolumn{2}{|c|}{{Year 3 }}&
    \multicolumn{2}{c|}{Year 5}\\
    \hline
    {\textbf{Model}}&{Cosine}&{Constant}&{Cosine}&{Constant}&{Cosine}&{Constant} \\
    
    \hline
$\mathbf{\chi^2}$\textbf{/DOF} & 59/36 & 57/37  & 154/108 & 159/109  & 237/181 & 246/182  \\ 
    \textbf{GOF}  & 0.002 & 0.004  & 0.00035 & 0.0002 & 0.0004 & 0.0001 \\
    \cline{2-7}
    \textbf{$\Delta \chi^2$ ($p$-value)} & \multicolumn{2}{c|}{1.57(0.2)} & \multicolumn{2}{c|}{-4.75 (0.029) } & \multicolumn{2}{c|}{-9.5 (0.002)} \\
    
\textbf{$Z$-score} & \multicolumn{2}{c|}{0.81$\sigma$} & \multicolumn{2}{c|}{1.89$\sigma$} & \multicolumn{2}{c|}{2.9$\sigma$} \\
    \hline
    \textbf{AIC(const) - AIC(cos)} & \multicolumn{2}{c|}{-2.57} & \multicolumn{2}{c|}{3.75} & \multicolumn{2}{c|}{8.5} \\
    \hline
    \textbf{BIC(const) - BIC(cos)} & \multicolumn{2}{c|}{-5.21} & \multicolumn{2}{c|}{0.05} & \multicolumn{2}{c|}{4.3} \\
    \hline
     \textbf{Bayes factor} & \multicolumn{2}{c|}{0.29} & \multicolumn{2}{c|}{5.4} & \multicolumn{2}{c|}{41} \\
     \hline
\end{tabular}
\caption{\rthis{Summary of model comparison tests on simulated data of a hypothetical dark matter experiment (similar to COSINE-100 or ANAIS-112 designed to detect annual modulation) consisting of simulated backgrounds (described in Table~\ref{table:simnoise}) and a simulated cosine  signal with amplitude $A=0.01$ cpd/kg/keV, equivalent to a  SNR of 0.29, where SNR is defined as the ratio of the amplitude of signal to the scale parameter of the simulated background. We find that after five years, the Bayes factor would provide decisive evidence for the cosine signal.}}
\label{table:simsignal}
\end{center}
\end{table*}

\begin{table*}
\begin{center}
\begin{tabular}{|c|c|c|c|c|c|c|}
    \hline
    {} & \multicolumn{2}{c}{{Year 1}} & \multicolumn{2}{|c|}{{Year 3 }}&
    \multicolumn{2}{c|}{Year 5}\\
    \hline
    {\textbf{Model}}&{Cosine}&{Constant}&{Cosine}&{Constant}&{Cosine}&{Constant} \\
    
    \hline
$\mathbf{\chi^2}$\textbf{/DOF} & 58/36 & 61/37  & 155/108 & 177/109  & 237/181 & 276/182  \\ 
    \textbf{GOF}  & 0.002 & 0.001  & 0.0003 & $8.9\times 10^{-6}$ & 0.0004 & $1.5\times 10^{-6}$ \\
    \cline{2-7}
    \textbf{$\Delta \chi^2$ ($p$-value)} & \multicolumn{2}{c|}{-2.8(0.09)} & \multicolumn{2}{c|}{-21 ($3 \times 10^{-6}$) } & \multicolumn{2}{c|}{-39 ($4.4 \times 10^{-10}$)} \\
    
\textbf{$Z$-score} & \multicolumn{2}{c|}{1.3$\sigma$} & \multicolumn{2}{c|}{4.5$\sigma$} & \multicolumn{2}{c|}{6.1$\sigma$} \\
    \hline
    \textbf{AIC(const)-AIC(cos)} & \multicolumn{2}{c|}{1.8} & \multicolumn{2}{c|}{21} & \multicolumn{2}{c|}{38} \\
    \hline
    \textbf{BIC(const)-BIC(cos)} & \multicolumn{2}{c|}{-0.8} & \multicolumn{2}{c|}{17} & \multicolumn{2}{c|}{34} \\
    \hline
     \textbf{Bayes factor} & \multicolumn{2}{c|}{1.7} & \multicolumn{2}{c|}{12731} & \multicolumn{2}{c|}{53272936} \\
     \hline
\end{tabular}
\end{center}
\caption{Summary of model comparison tests on simulated data of a hypothetical dark matter experiment (similar to COSINE-100 or ANAIS-112 designed to detect annual modulation) consisting of simulated backgrounds (described in Table~\ref{table:simnoise}) and  simulated cosine  signal with amplitude $A=0.02$ cpd/kg/keV, equivalent to  a SNR (defined in Table~\ref{table:simsignal}) of 0.6.  We find that after three years of exposure, both the information theory tests (AIC and BIC) and Bayes factor would provide decisive evidence for the cosine signal.}
\label{table:simsignal2}
\end{table*}

\subsection{Results with simulated cosine signal}
\rthis{The DAMA best-fit value for the amplitude of the cosine signal is $\sim 0.01$ cpd/kg/keV.
The actual detectability of this signal in another dark matter experiment would depend upon the signal to noise ratio (or the ratio of amplitude of signal to  that of the background). In lieu of more realistic background models for the residual rates (which can be done by the experimental collaborations), we added simulated signals with two different amplitudes to the simulated backgrounds (described in previous sub-section) in order to assess the significance with different signal to noise ratios. Our simulated cosine signal has a period of one year and a phase of 159 days, identical to the best-fit values found by the DAMA collaboration. The values of the injected  amplitudes are 0.01 cpd/kg/keV and 0.02 cpd/kg/keV. For the background scale parameter of 0.035 cpd/kg/keV, this corresponds to a SNR of 0.28 and 0.57 respectively. We note that the results with the SNR of 0.57 would be identical for a signal with amplitude of 0.01 cpd/kg/keV and with a background scale parameter half of our currently simulated value. 
We note that for doing the fits (similar to the analysis of data with only noise), the period of the cosine signal was kept fixed at one year. Only the amplitude and phase were allowed to vary.
Our results for SNRs of 0.28 and 0.57 can be found in Tables~\ref{table:simsignal} and \ref{table:simsignal2}.}

\rthis{We note that for the  SNR of 0.28~(Tab.~\ref{table:simsignal}), after one year of data, the different model comparison tests marginally prefer a constant model. After three years of data, all the tests prefer the cosine model. However, only the Bayes factor provides substantial evidence for the cosine model according  to the Jeffreys scale~\cite{Trotta}.
After five years of data, both AIC and Bayes factor provide strong evidence for the cosine model.}

\rthis{For the signals with SNR of 0.57 (Tab.~\ref{table:simsignal2}), after the first year of data, except BIC, all the other tests very marginally prefer the cosine model. After three years of accumulated exposure, the information theory based tests and Bayes factor show decisive evidence for a cosine signal. The $Z$-score  for the frequentist test after three years is 4.5$\sigma$ and increases to 6$\sigma$ only after five years. Therefore, we find that the Bayes factor is most sensitive in detecting the cosine signal even for weak amplitudes.}

\section{Conclusions}
\label{sec:conclusions}
The DAMA  experiment, which looks for direct detection of dark matter, has found evidence for annual modulation for more than 20 years, with all the right characteristics to be caused by galactic dark matter scattering. In their latest data release paper from phase 2 of their upgraded experiment (called DAMA/LIBRA)~\citep{Dama18}, the statistical significance ranges from  $8\sigma$ and $12.9\sigma$, depending on the  energy  intervals and durations~\citep{Dama18}.

This statistical significance has been evaluated using a frequentist technique, which involves comparing the difference in $\chi^2$ between the null hypothesis and sinusoidal model. From this difference, a $p$-value can calculated using Wilk's theorem, which is then converted to a $Z$-score or significance, in terms of number of sigmas.

In this work, we independently assess the DAMA claim for annual modulation using the same frequentist model comparison technique, by including the errors in the independent time-variable. Furthermore, we use model comparison tests from information theory and Bayesian analysis, which are routinely used in Cosmology. For the information theory tests, we used the Akaike Information Criterion and Bayesian Information Criterion to evaluate the significance. From the difference, the significance was evaluated using qualitative strength of evidence rules.
For Bayesian model comparison, we calculated the marginal likelihood or Bayesian evidence for both the hypotheses. For this purpose, we assumed uniform priors on all the models. We then calculated the Bayes factor from the ratio of the marginal likelihood and used Jeffrey's scale to evaluate the significance. 

A tabular summary of all our model comparison tests can be found in Table~\ref{table:5}. The Bayesian and information theoretical comparison tests also point to very strong evidence for annual modulation and agree with the frequentist tests.

\rthis{We also performed a search for higher harmonics in the DAMA data by looking at the residual signal, obtained by subtracting the best-fit annual modulation signal from the  raw residual background. We applied the generalized Lomb-Scargle periodogram~\cite{Kurster,vanderplas} to this residual data and found another peak in both 1-6 keV and 2-6 keV data at a period of 40 days. However, the significance is very low ($<3 \sigma$) and this period is close to the DAMA bin size of around 20 days. So more data is necessary to ascertain if the observed significance of this peak increases.}

\rthis{Finally, we also carried out numerical experiments to check how robustly the different model comparison techniques would help discriminate between the two hypothesis for a dark matter experiment sensitive to annual modulation, using a simulated data stream consisting of pure background as well as a superposition of pure background and cosine signal with different amplitudes. The simulated background has been designed to mimic the residual background of the recently released ANAIS-112 data~\cite{Anais}. We find that if the data contained only background, none of the model comparison techniques used can decisively select any one model (cf. Table~\ref{table:simnoise}). When we injected signals with two different amplitudes, corresponding to SNRs of 0.28 and 0.57~(cf. Tables~\ref{table:simsignal} and ~\ref{table:simsignal2}), we find that the Bayes factor is the most sensitive and can provide decisive evidence for the cosine signal even for SNR of 0.28 after five years of exposure and after three years for SNR of 0.56}.

This is the first proof of principles application of information theoretic and Bayesian model comparison techniques to evaluate the annual modulation claim in DAMA
\rthis{and to analyze the sensitivity of other dark matter experiments such as ANAIS-112 and COSINE-100 to look for such signals}.

All our analysis codes and data to reproduce these results are available online at \url{https://github.com/aditikrishak/DAMA_Model_Comparison}.

\begin{acknowledgements}
Aditi Krishak is supported by a DST-INSPIRE fellowship. 
\end{acknowledgements}

\newpage
\bibliography{main}
\end{document}